\documentclass[aps,twocolumn,prc,floatfix,showpacs,preprintnumbers,amsmath,amssymb, nofootinbib,groupedaddress]{revtex4-1}

\usepackage{amsmath,amssymb,amsthm,amsfonts}
\usepackage{graphicx}
\usepackage{color}
\usepackage{ulem}
\usepackage{bm} 

\begin{document}

\title{Regularization of zero-range effective interactions in finite nuclei}

\author{Marco Brenna}
\affiliation{Dipartimento di Fisica, Universit\`a degli Studi di Milano and INFN,  Sezione di Milano, 20133 Milano, Italy}

\author{Gianluca Col\`o}
\email{Gianluca.Colo@mi.infn.it}
\affiliation{Dipartimento di Fisica, Universit\`a degli Studi di Milano and INFN,  Sezione di Milano, 20133 Milano, Italy}

\author{Xavier Roca-Maza}
\affiliation{Dipartimento di Fisica, Universit\`a degli Studi di Milano and INFN,  Sezione di Milano, 20133 Milano, Italy}

\date{\today} 

%
\begin{abstract}
The problem of the divergences which arise in beyond mean-field 
calculations, when a zero-range effective interaction is employed, 
has not been much considered so far. Some of us have proposed, quite 
recently, a scheme to regularize a zero-range Skyrme-type force when 
it is employed to calculate the total energy, at second-order 
perturbation theory level, in uniform matter. Although this 
scheme looked promising, the extension for finite nuclei is 
not straightforward. We introduce such procedure in the current paper, by proposing a regularization procedure that is similar, in spirit, to the one employed to extract the so-called $V_{\rm low-k}$ from the bare force. Although this has been suggested already by B.G. Carlsson and collaborators, the novelty of our work consists in setting on equal footing uniform matter and finite nuclei; in particular, we show how the interactions that have been regularized in uniform matter behave when they are used in a finite nucleus with the corresponding cutoff. We also address the problem of the validity of the perturbative approach in finite nuclei for the total energy.
\end{abstract}   

%


%

\maketitle

%
\section{Introduction}

Self-consistent mean-field approaches provide a fairly good starting
approximation to describe atomic nuclei \cite{Bender}. Whereas so-called 
{\it ab-initio} approaches are increasingly successful, they cannot
at present describe heavy systems and/or high-lying excited states.
Mean-field approaches, instead, are able to
reproduce both the experimentally observed trends of many ground-state
properties (masses, radii, deformations etc.) and several features of excited states
(giant resonances, rotational bands); moreover, they can be extended if needed.
Either non relativistic Hamiltonians of the Skyrme or Gogny type, or
covariant relativistic mean-field (RMF) Lagrangians have been used indeed beyond the mean-field
approximation, for instance in second-order calculations \cite{VanNeck1,VanNeck2}, in multiparticle-multihole 
schemes \cite{Pillet}, in particle-vibration coupling (PVC) models \cite{Bernard,Colo:2010,Mizuyama,Ogata,Colo:1992,
Colo:1994,Niu,Litvinova1,Afanasjev1,Afanasjev2,LitvinovaGR1a,*LitvinovaGR1b,LitvinovaGR2,LitvinovaGR3}, or within the 
generator coordinate method (GCM)  approach \cite{GCM1,GCM2,GCM3,GCM4}. 

In such approaches one introduces further correlations on top
of those implicit in the mean field. Within PVC, the
nucleons feel the effect of the dynamical fluctuations of the mean field,
on top of its static part; within GCM, the variational space associated with a 
single mean-field 
configuration
is enlarged by superimposing several mean-field configurations, each being connected with 
a different value of some global parameter like the quadrupole deformation.
If effective interactions are fitted at mean-field level one would
imagine that a re-fit of these interactions is mandatory if they
are employed in a different framework. However, this
is usually not done and Skyrme and Gogny forces or RMF Lagrangians are used
as they are. We have in mind, for the follow-up of our discussion, mainly the PVC case since we
shall consider in detail the lowest-order approximation to that model.

If single-particle (s.p.) nuclear states are calculated using Skyrme
or RMF Lagrangians, the corrections induced by PVC at lowest
order are typically several hundreds of keV, ranging from small
values to $\approx$ 1-2 MeV \cite{JPhysG}. These corrections improve
as a rule the agreement with experiment, leading to a r.m.s. deviation
with respect to experimental values of about 1 MeV or less in e.g. 
$^{208}$Pb \cite{Colo:2010,Ligang:tbp}. However, the convergence
of these results with respect to the model space is hard to assess.
Normally, one assumes that the model space is limited by the fact
that only collective vibrations should be taken into account but
this does not set a cutoff in a clear-cut way.

At the same time, there is another practical and yet more general point to be kept in
mind. When zero-range forces are used in a beyond mean-field
approach like PVC, divergences arise. In other words, diagrams
beyond the HF ones (like those displayed in Fig. 1 of \cite{Colo:2010}) can be shown, by
simple power counting arguments, to diverge as the model space
is enlarged. This is not a specific problem of Skyrme as also
Gogny forces possess a zero-range term. We do not dispose at present
of a reliable, fully microscopic non relativistic Hamiltonian without zero-range
terms. As a consequence, it is necessary to devise a regularization
technique to absorb this divergence and go beyond the usual PVC
calculations. In the current work, we focus on the Skyrme case. 

To simplify the formidable problem of finding a reliable regularization technique for
nuclei, we only consider in the following the lowest-order (that is, second order) approximation for
PVC in which the phonon is replaced by a particle-hole (p-h) pair, and we
focus on the correction to the total energy instead of the correction to the
s.p. energies. This problem has been already tackled in uniform matter 
\cite{Mog:1,Mog:2}, where it has been shown that at least a cutoff
regularization is possible for a general Skyrme interaction and an arbitrary
neutron/proton ratio. A dimensional regularization technique 
\cite{Mog:dim} has also been proposed, whereas the general study of the 
renormalizability of a Skyrme-type force has been recently 
addressed in Ref. \cite{grasso13}. Also these studies concern only
uniform matter. The extension of the techniques introduced in 
Refs. \cite{Mog:1,Mog:2} to finite nuclei is far from being 
straightforward and is the subject of the current paper.

\begin{figure}[t]
\includegraphics[width=0.8\linewidth]{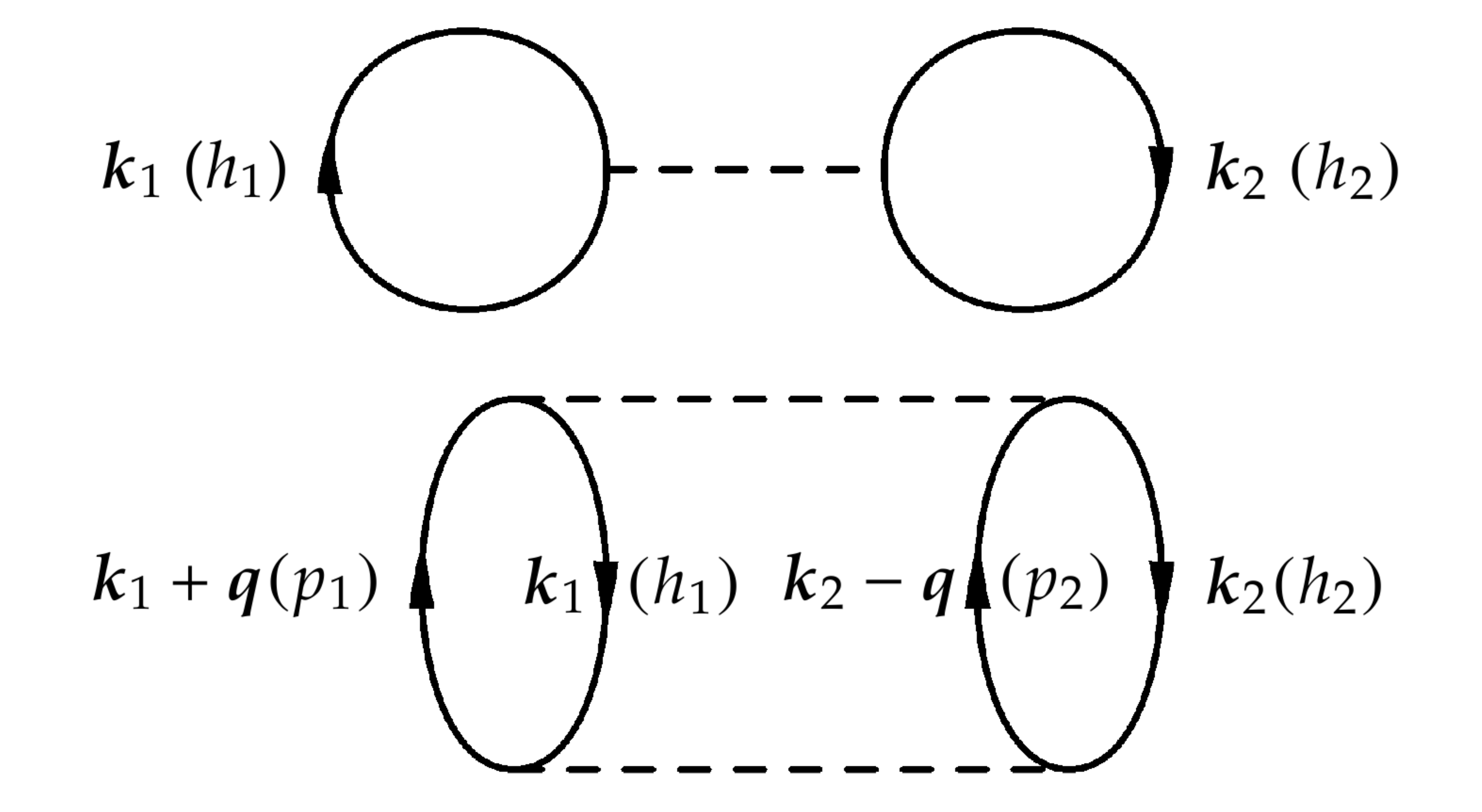}
\caption{Diagrammatic representation of the first-order (Hartree-Fock) and second-order total
energy, respectively in the upper and lower parts of the figure. The labels outside (inside)
parenthesis are those used in the text for the case of uniform matter (finite nuclei).}
\label{fig00}
\end{figure}

The total energy is depicted diagrammatically in Fig. \ref{fig00}. We have drawn only
the direct contributions but exchange terms are properly included in our
calculations. The first row of the figure depicts the mean-field 
or Hartree-Fock (HF) total energy and the second line the second-order
contribution to the same quantity. The labels outside (inside) 
parenthesis are momentum (generic) labels, appropriate for uniform matter 
(nuclei) respectively.

\begin{figure}[t]
\includegraphics[width=\linewidth]{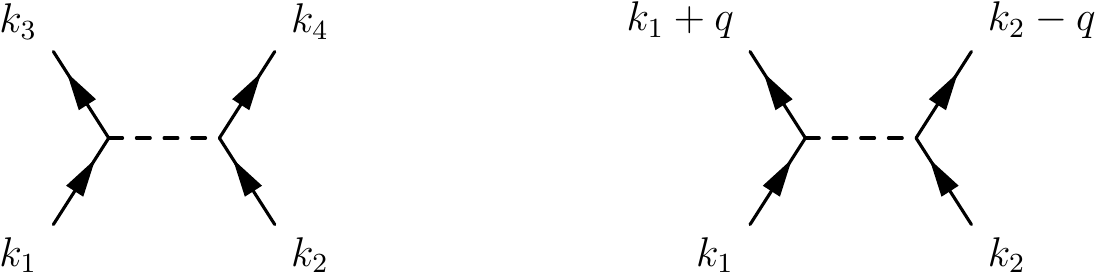}
\caption{Representation of matrix elements in the case of uniform matter. 
We compare the notation used in the present paper (left) with 
the one in which the transferred momentum $q$ appears (right).}
\label{fig0}
\end{figure}

In the case of uniform matter, a simple power counting argument dictates
that the second order contribution diverges. The simplest way to understand it is the
following. The integration on momentum states is finite with
respect to the hole momentum states $k_1$, $k_2$ which
have the Fermi momentum $k_F$ as an upper limit. The center-of-mass momentum
conservation leaves only one further momentum scale, which has been
chosen in Refs. \cite{Mog:1,Mog:2} to be the transferred momentum. 
If we label the particle states needed for the calculation of the second-order energy as
$k_3$, $k_4$, the matrix elements are     
\begin{equation}\label{me}
\langle \bm{k_3}, \bm{k_4} \vert V \vert \bm{k_1}, \bm{k_2} \rangle  
=
\langle \bm{k_1}+ \bm{q}, \bm{k_2} - \bm{q} \vert V \vert \bm{k_1}, \bm{k_2} \rangle  
\end{equation}
(cf. Fig. \ref{fig0}), and the quantity
\begin{equation}
\bm{q} \equiv \frac{\bm{k_3} - \bm{k_4} - \bm{k_1} + \bm{k_2} }{2}
\label{eq:q}
\end{equation} 
is the transferred momentum. For a zero-range interaction without velocity dependence (i.e., a
pure $\delta$-force) the second-order contribution
diverges linearly, or in other words it scales as $\int \frac{d^3q}{q^2}$, 
while the divergence is more severe if momentum-dependent 
terms are included.
In Refs. \cite{Mog:1,Mog:2} 
it has been shown that, by setting a cutoff $\Lambda$ on the transferred momentum, 
given an interaction $V$ that provides a reasonable total energy at mean-field level, 
it is possible to fit an interaction $V_\Lambda$ that
reproduces the same energy when the second-order correction
has been included with the cutoff $\Lambda$. 

In finite nuclei, at variance with uniform matter, there is not 
translational invariance and in dealing with the matrix elements 
(\ref{me}) we are left with two free parameters or two energy scales.
In the current work we have dealt with
the two scales by defining in a precise fashion the relative and center-of-mass
coordinates and the associated momenta. 
Since the separation of center-of-mass and relative motion 
wave functions can be done 
in a neat way by using an harmonic oscillator basis, the calculations 
that we shall discuss below have been performed on that
basis. We have systematically defined a cutoff $\lambda$ on the relative momenta
(in initial and final channel) defined as
\begin{eqnarray}
\bm{k} & \equiv & \frac{\bm{k_1}-\bm{k_2}}{\sqrt{2}}, \nonumber \\
\bm{k'} & \equiv & \frac{\bm{k_3}-\bm{k_4}}{\sqrt{2}}.
\label{eq:firstkk}
\end{eqnarray}

Then, in our study, we will show how a simplified Skyrme interaction, 
in which
only the $t_0$, $t_3$ and $\alpha$ parameters are kept and which has 
been regularized in uniform matter, behaves when it is used in a finite 
nucleus. We restrict ourselves to the case of even-even, 
isospin-symmetric nuclei; in particular, we will show results 
for $^{16}$O without Coulomb and spin-orbit forces. Our approach 
is thus self-consistent in the sense that we use the same Skyrme 
interaction both at mean-field and second order level, 
but we compute the total energy at second order in a perturbative way, 
by adding the beyond mean-field contribution on top of HF solutions. 
We will eventually address the problem of the validity of such perturbative 
approach for the total energy of a finite nucleus.

The structure of the paper is the following. Sec. \ref{forma} is devoted to a thorough explanation
of the formalism we wish to introduce: in particular, in subsec. \ref{2a} we 
discuss the interaction and its regularization, whereas in the next subsection \ref{2b} we show the relationship 
between the new cutoff employed in this work 
and the cutoff that had been introduced previously in uniform matter. The specific formulas
that implement the regularized calculation of the total energy in a finite nucleus, on the harmonic oscillator basis, 
are introduced in subsec. \ref{2c}. In Sec. \ref{res} we describe the results obtained in the case of 
$^{16}$O. Conclusions and considerations related to the envisaged follow-up of this work, are 
in Sec. \ref{concl}. 

%

\section{Formalism}\label{forma}

\subsection{The regularization of the interaction}\label{2a}

The cutoff on the relative momentum components of the effective interaction is analogous to that discussed
in Ref.~\cite{carls13}. The underlying philosophy is the same as in the case of the $V_{\rm low-k}$ interaction and it seems quite natural,
even by invoking the original argument that the Skyrme interaction is a polynomial expansion in the relative momentum that stops 
at second order \cite{skyrme59a}. Therefore, we introduce the cutoffs on the relative momenta of the
initial and final states and we define a regularized interaction through these cutoffs.
In principle, this procedure could be avoided by using a finite-range force. However, as we stressed
in the Introduction, we miss at present a widely used, reliable microscopic pure finite-range force. 

To identify properly the relative momenta we introduce center-or-mass and relative coordinates. 
We start by writing the velocity-independent part of the Skyrme force in this form:
\begin{equation}
V(\bm{r}'_1,\bm{r}'_2,\bm{r}^{}_1,\bm{r}^{}_2) =
g \left( \frac{\bm{r}_1+\bm{r}_2}{2} \right)
\delta (\bm{r}_1-\bm{r}_2)
\delta (\bm{r}_1-\bm{r}'_1)
\delta (\bm{r}_2-\bm{r}'_2).
\label{eq:interr1r2}
\end{equation}
Our desired change of variables reads
\begin{equation}
\left( \begin{array}{l}
\bm{r}^{(\prime)} \\
\bm{R}^{(\prime)}
\end{array} \right)
= \left(
\begin{array}{rr}
\frac{1}{\sqrt{2}} & -\frac{1}{\sqrt{2}} \\
\frac{1}{\sqrt{2}} & \frac{1}{\sqrt{2}} \\
\end{array} \right) \left(
\begin{array}{l}
\bm{r}^{(\prime)}_1 \\
\bm{r}^{(\prime)}_2
\end{array} \right), 
\label{eq:transIordrR}
\end{equation}
so that the interaction \eqref{eq:interr1r2} can be written as
\begin{equation}
\begin{split}
V(\bm{r}',\bm{R}',\bm{r}^{},\bm{R}^{}) & =
\frac{\sqrt{2}}{4} g \biggl( \frac{\bm{R}}{\sqrt{2}} \biggr)
\delta(\bm{r})
\delta(\bm{r}')
\delta(\bm{R}-\bm{R}') \\
& =
\frac{\sqrt{2}}{4} g \biggl( \frac{\bm{R}}{\sqrt{2}} \biggr)
v(\bm{r}',\bm{r})
\delta(\bm{R}-\bm{R}'),
\end{split}
\label{eq:interaction}
\end{equation}
where $g(\bm{R}) = t_0 + \frac{t_3}{6}\left[\rho \left(\bm{R}\right) \right]^\alpha $.
The Fourier transform of the interaction can be written in a straightforward way as
\begin{widetext}
\begin{equation}
\begin{split}
V\left(\bm{k}_3,\bm{k}_4,\bm{k}_1,\bm{k}_2 \right) =  
\frac{\sqrt{2}}{4} &\frac{1}{\Omega} \int \mathrm{d}^3 R^{} \mathrm{d}^3 R' \,
\mathrm{e}^{-i \frac{\bm{k}_3 + \bm{k}_4}{\sqrt{2}} \! \cdot \! \bm{R}'}
g \biggl( \frac{\bm{R}}{\sqrt{2}} \biggr) \delta \left(\bm{R}-\bm{R}'\right)
\mathrm{e}^{i \frac{\bm{k}_1 + \bm{k}_2}{\sqrt{2}} \! \cdot \! \bm{R}^{}} \times \\
\times & \frac{1}{\Omega}
\int \mathrm{d}^3 r \mathrm{d}^3 r' \,
\mathrm{e}^{-i \frac{\bm{k}_3 - \bm{k}_4}{\sqrt{2}} \! \cdot \! \bm{r}'}
v(\bm{r}',\bm{r})
\mathrm{e}^{i \frac{\bm{k}_1 - \bm{k}_2}{\sqrt{2}} \! \cdot \! \bm{r}},
\end{split}\label{ft0}
\end{equation}
\end{widetext}
by introducing a finite quantization volume $\Omega$. 
The factor appearing in the second line of this equation can be written in terms of the variables 
$\bm{k} \equiv \frac{\bm{k}_1 - \bm{k}_2}{\sqrt{2}}$ and $\bm{k}' \equiv \frac{\bm{k}_3 - \bm{k}_4}{\sqrt{2}}$
that are the conjugate variables of the relative coordinates defined by Eq. (\ref{eq:transIordrR}): these are
the relative momenta (of the initial and
final state respectively) that we have already introduced in Eq. (\ref{eq:firstkk}). Thus we re-write the 
factor in the second line of Eq. (\ref{ft0}) as 
\begin{align}
v(\bm{k}',\bm{k}) &= \frac{1}{\Omega} \int \mathrm{d}^3 r \mathrm{d}^3 r' \, \mathrm{e}^{-i \bm{k}' \! \cdot
\bm{r}'} v (\bm{r}', \bm{r} ) \mathrm{e}^{i \bm{k} \! \cdot \bm{r}} 
= \frac{1}{\Omega}, \label{eq:Fourvk}
\end{align}
where the last equality obviously holds if
$v(\bm{r}',\bm{r}) = \delta(\bm{r})\delta(\bm{r}')$ as
we have written in Eq. \eqref{eq:interaction} (we will keep this
notation in what follows).

Then, we introduce the regularized interaction as the inverse Fourier transform of \eqref{eq:Fourvk} in which two step functions 
$\theta(\lambda - k) \theta(\lambda' - k')$ are introduced. In this way, $\lambda$ and $\lambda^\prime$ are the cutoffs in the relative
momenta $\bm{k}$ and $\bm{k'}$, respectively, and the regularized interaction $v^{\lambda\lambda'}$ is obtained as 
\begin{widetext}
\begin{align}
v^{\lambda \lambda'}(\bm{r}', \bm{r}) = & \frac{1}{\Omega} \int \mathrm{d}^3 k \mathrm{d}^3 k' \, \mathrm{e}^{i \bm{k}' \!
\cdot \bm{r}'} v\left(\bm{k}', \bm{k}\right) \theta(\lambda - k) \theta(\lambda' - k') \mathrm{e}^{-i \bm{k} \! \cdot \bm{r}} \nonumber \\
= & \frac{1}{4 \pi^4} \frac{\lambda^2 \lambda'^2}{r r'} j_{1}(r \lambda) j_{1}(r' \lambda') \label{eq:renint}
\xrightarrow[\substack{\lambda \rightarrow +\infty \\ \lambda' \rightarrow +\infty}]{} 
\delta(\bm{r}) \delta(\bm{r}'),
\end{align}
\end{widetext}
where the usual expansion of the plane waves in spherical components is used, and the limit in the last line 
comes from Eq.~(3.5) of Ref.~\cite{mehr11}.  

In what follows, we will employ the regularized interaction $v^{\lambda \lambda'}(\bm{r}', \bm{r})$ 
to evaluate the matrix elements of the interaction \eqref{eq:interaction}, and at times compare with the
matrix elements obtained by using the bare interaction $v(\bm{r}', \bm{r})$. 

\subsection{Uniform matter and the different choices for the cutoff}\label{2b}

In this subsection, we wish to establish a connection between the cutoff on the transferred momentum (\ref{eq:q}) \cite{Mog:1,Mog:2} 
and the cutoff on the relative momenta (\ref{eq:firstkk}). At the same time, we deal in this
subsection with that fact that in the procedure adopted in Refs. \cite{Mog:1,Mog:2} there 
is no cutoff affecting the HF energy. 
In the present scheme, we introduce a cutoff consistently in the HF and second-order energies.

\begin{figure}
\centering
\includegraphics[width=0.9\linewidth]{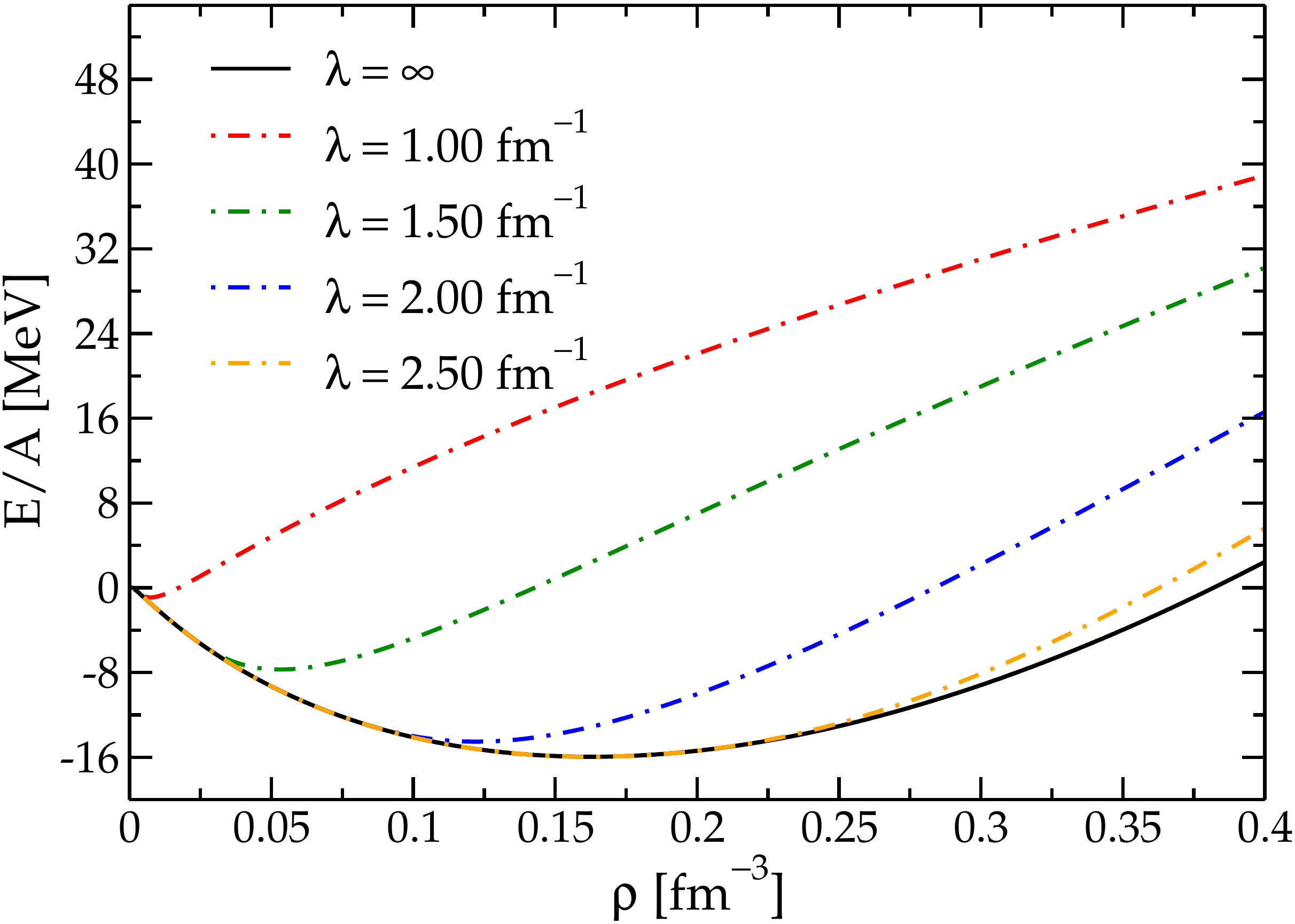}
\caption{(Color online) Energy per particle at the HF level [cf. Eqs. (\ref{eq:HF},\ref{eq:Iordcut})] for different values of the cutoff $\lambda$ 
on the relative momentum $\bm{k}$. \label{fig:eos}}
\end{figure}

The HF 
potential energy, shown diagrammatically in the upper part of
Fig. 1, is 
\begin{equation}
E_{\rm HF} = \frac{1}{2}\sum_{ij} \langle ij \vert \bar{V} \vert ij \rangle,
\label{eq:HF}
\end{equation}
where $\bar{V}$ is the antisymmetrized interaction. If we write the HF energy 
in symmetric nuclear matter as in Ref. \cite{Mog:1}, 
we obtain
\begin{equation}
\frac{E}{A} = 8 \frac{d g k_F^6}{\rho (2\pi)^6} \frac{4 \pi}{3} \int \mathrm{d}^3 \tilde{k} \,
\left(1-\frac{3}{2} \tilde{k}+\frac{1}{2} \tilde{k}^3\right) \theta (1-\tilde{k}) =
\frac{3}{8}g \rho,
\label{eq:Iordnocut}
\end{equation}
where $d$ is the level degeneracy (4 in the case of symmetric nuclear 
matter) and $\tilde {\bm k}$ is defined only in this subsection,
for the sake of convenience, as
$\tilde {\bm k} \equiv {\bm k} / \sqrt{2}k_F$. If we now wish to introduce the 
cutoff $\lambda$ on $\bm{k}$, we have to add a factor
$\theta \! \left(\frac{\lambda}{\sqrt{2} k_F}-\tilde{k}\right)$. Then, Eq.~\eqref{eq:Iordnocut} becomes
\begin{widetext}
\begin{equation}
\frac{E}{A} = 8\frac{d g k_F^6}{\rho (2\pi)^6} \frac{4 \pi}{3} \int \mathrm{d}^3 \tilde{k} \,
\left(1-\frac{3}{2}\tilde{k}+\frac{1}{2}\tilde{k}^3\right)\theta (1-\tilde{k}) \theta \! \left(\frac{\lambda}{\sqrt{2} k_F}-\tilde{k}\right) = \frac{3}{8}g \rho \left(8\beta^3 -9 \beta^4 +2 \beta^6\right),
\label{eq:Iordcut}
\end{equation}
\end{widetext}
where $\beta = \min\{1,\frac{\lambda}{\sqrt{2} k_F}\}$. Clearly, if $\lambda > \sqrt{2} k_F$, then $\beta = 1$ and we recover the
result of Eq.~\eqref{eq:Iordnocut}. This has been tested also numerically, and the result is displayed in Fig. \ref{fig:eos}. Note
the similar figure and reasoning in Ref. \cite{carls13}.

As for the second order contribution to the total energy, the relation between momenta used in the present work
and those employed previously \cite{Mog:1,Mog:2} can be written,
generalizing Eq. (\ref{eq:firstkk}), as
\begin{equation}
\left( \begin{array}{l}
\bm{k} \\ \bm{k}' \\ \bm{k}'' \end{array} \right)
= \left( \begin{array}{rrr}
\frac{1}{\sqrt{2}} & -\frac{1}{\sqrt{2}} & 0 \\
\frac{1}{\sqrt{2}} & -\frac{1}{\sqrt{2}} & \sqrt{2} \\
\frac{1}{\sqrt{2}} & \frac{1}{\sqrt{2}} & 0 \\ \end{array} \right) \left(
\begin{array}{l}
\bm{k}_1 \\ \bm{k}_2 \\ \bm{q} \end{array} \right).
\end{equation}
The determinant of the Jacobian matrix of this transformation is 1. 
One can note that 
\begin{equation}
{\bm k'} = {\bm k} + \sqrt{2}{\bm q}.
\end{equation} 
The second order contribution to the total energy, displayed in the lower part of Fig. 1 with diagrams, is
\begin{equation}\label{secondorder1}
\Delta E = \frac{1}{2} \sum_{ijmn}\frac{\langle mn \vert V \vert ij \rangle \langle ij \vert \bar{V} \vert mn \rangle}{\varepsilon_i
+\varepsilon_j-\varepsilon_m-\varepsilon_n},
\end{equation}
where $\varepsilon$ are HF s.p. energies. 
We evaluate this expression in symmetric matter and 
we keep the notation of this subsection, that is, $\tilde {\bm k} \equiv
{\bm k} / \sqrt{2}k_F$ and 
$\tilde {\bm k}' \equiv {\bm k}' / \sqrt{2}k_F$, 
$\tilde {\bm k}'' \equiv {\bm k}'' / \sqrt{2}k_F$
in an analogous way. Thus, we obtain
\begin{equation}
\frac{\Delta E}{A} = \chi(\rho) \frac{\sqrt{2}}{4 \pi^3}
\int_{\mathcal{D}(\tilde {\bm{k}},\tilde {\bm{k}}',\tilde {\bm{k}}'')}
\mathrm{d}^3 \tilde {k} \mathrm{d}^3 \tilde {k}' \mathrm{d}^3 \tilde {k}'' \, 
\frac{1}{\tilde {k}'^2 - \tilde {k}^2},
\label{eq:IIordcarl}
\end{equation}
where $\chi(\rho)$ has been defined in Ref. \cite{Mog:1}, and the domain of integration is
\begin{widetext}
\begin{multline*}
\mathcal{D}(\tilde {\bm{k}},\tilde {\bm{k}}',\tilde {\bm{k}}'') \equiv 
\biggl\{\tilde {\bm{k}},\tilde {\bm{k}}',\tilde {\bm{k}}'' \in 
\mathbb{R}^3\colon \tilde {k} \le 1,
\tilde {k}''\le 1,
\left( \left| \tilde {\bm{k}}'' + \tilde {\bm{k}} \right| < 1 \cap \left| 
\tilde {\bm{k}}'' - \tilde {\bm{k}} \right| < 1 \right) \cup
\left( \left| \tilde {\bm{k}}'' + \tilde {\bm{k}}' \right| > 1 \cap \left| 
\tilde {\bm{k}}'' - \tilde {\bm{k}}' \right| > 1 \right) \biggr\}.
\end{multline*}
\end{widetext}

Our purpose is now to compare with the results of Ref.~\cite{Mog:1} 
and convince ourselves
that we can use the interactions that have been fitted therein. 
To this aim, we must consider the
case in which the cutoff $\lambda$ on relative momenta is 
larger than $\sqrt{2}k_F$, since otherwise the
HF energy should be also modified compared 
to the calculation with the bare force performed in Ref.~\cite{Mog:1}
(cf. above). On top of this, the calculation of the integral appearing in Eq. \eqref{eq:IIordcarl} is rather
cumbersome, and can be slightly simplified if $\lambda$ is larger than $2\sqrt{2}k_F$. In this case a
detailed analytic evaluation has been carried out \cite{brenna}. The result can be written as
\begin{widetext}
\begin{equation}
\begin{split}
\frac{\Delta E}{A} = \chi(\rho) \biggl\{&-\frac{11}{105}+
\frac{2}{105}\ln{2}+\frac{2}{35}\tilde{\lambda} -
\frac{11}{35} \tilde{\lambda}^3
- \frac{2}{21} \tilde{\lambda}^5
-\left( 
\frac{4\tilde{\lambda}^5}{5} - \frac{4 \tilde{\lambda}^7}{21} \right) 
\ln\left(\tilde{\lambda}\right)\\
&+\left(\frac{1}{35}-
\frac{\tilde{\lambda}^4}{3}+\frac{2 \tilde{\lambda}^5}{5}-
\frac{2 \tilde{\lambda}^7}{21} \right) \ln\left(\tilde{\lambda}-1 \right)
-\left(\frac{1}{35}- \frac{\tilde{\lambda}^4}{3}
-\frac{2 \tilde{\lambda}^5}{5}
+\frac{2 \tilde{\lambda}^7}{21} \right) \ln\left(\tilde{\lambda}+1 \right)
\biggr\},
\end{split}
\label{eq:IIordcarlfin}
\end{equation}
\end{widetext}
where $\tilde{\lambda} \equiv \frac{\lambda}{\sqrt{2}k_F}$. 
We have checked that the part that does not depend on 
$\lambda$ is equal to the one already written in Ref.~\cite{Mog:1}, as it should, and that the divergence is linear.

To obtain a better understanding, we have evaluated numerically the two results given by 
Eq.~\eqref{eq:IIordcarlfin} and Eq.~(8) of Ref.~\cite{Mog:1}. 
The two calculations are almost indistinguishable when 
\begin{equation}\label{twocutoffs}
\lambda = \sqrt{2} \Lambda.
\end{equation}
We will then use this latter equation in the following way: 
when we perform a calculation of a finite system with cutoff
$\lambda$ we will adopt the interaction fitted in uniform 
symmetric matter with the value of $\lambda$ given by
Eq.~(\ref{twocutoffs}). Ultimately, we would 
envisage to cast uniform matter and finite nuclei in a single scheme, 
so to be able to fit an effective force 
in the same spirit of the original Skyrme force 
(at second order and then beyond). 

\subsection{The formalism for finite nuclei using the harmonic oscillator basis}\label{2c}

In finite nuclei the second order energy is still given by Eq. 
(\ref{secondorder1}) but is more conveniently written in angular 
momentum-coupled representation as
\begin{equation}\label{de-finite-1}
\Delta E = \frac{1}{4} \sum_{pp'hh'J} \frac{(2J+1)\vert \langle (pp')J \vert 
\bar{V} \vert (hh')J \rangle \vert^2}{\varepsilon_h+\varepsilon_{h'}
-\varepsilon_p-\varepsilon_{p'}},
\end{equation}
where the 
particle-particle (pp) coupled matrix elements have been introduced,
namely
\begin{widetext}
\begin{align}
\begin{aligned}
\langle (\alpha \beta) J &M_J | \bar{V} | (\gamma \delta) J M_J \rangle =
\sum_{\substack{ m_{\alpha} m_{\beta} \\ m_{\gamma} m_{\delta}}} \langle j_{\alpha} m_{\alpha} j_{\beta} m_{\beta} | J M_J
\rangle \langle j_{\gamma} m_{\gamma} j_{\delta} m_{\delta} | J M_J \rangle \langle \alpha \beta | \bar{V} | \gamma \delta
\rangle\\
&=
\sum_{\substack{ m_{\alpha} m_{\beta} \\ m_{\gamma} m_{\delta}}}
(-)^{j_{\alpha}-j_{\beta}+j_{\gamma}-j_{\delta}} \hat{J}^2
\begin{pmatrix}
j_{\alpha} & j_{\beta} & J \\
m_{\alpha} & m_{\beta} &-M_J
\end{pmatrix}
\begin{pmatrix}
j_{\gamma} & j_{\delta} & J \\
m_{\gamma} & m_{\delta} &-M_J
\end{pmatrix}
\langle \alpha \beta | \bar{V} | \gamma \delta \rangle,
\end{aligned}
\label{eq:ppcouplme}
\end{align}
\end{widetext}
where we have introduced the common shorthand notation 
$\hat{J}^2 = 2J+1$. Actually, these latter matrix 
elements do not depend on $M_J$ because
of rotational invariance. Therefore, in Eq. (\ref{de-finite-1}) they
appear without this label; in that equation, the trivial sum
over $M_J$ has been performed. 

In our calculation, we expand the single-particle wave functions in harmonic oscillator basis. Then, the corresponding matrix elements 
are evaluated by performing the transformation 
of the initial and final two-particle
states to the center of mass and relative motion coordinates. As is well known, this can be done in the HO case by 
exploiting the Brody-Moshinsky transformations \cite{lawson,buck96,kamun01}. 
In this subsection we will collect only the main
equations related to the matrix elements entering our calculations; 
we shall provide some more details about the main steps of 
their derivation, together with other useful formulas, in the Appendix. 

We shall indicate with the label $I$ = 
$0$, $\sigma$, $\tau$ and $\sigma \tau$ the terms
of the pp-coupled matrix elements~\eqref{eq:ppcouplme} that
are proportional, respectively, to the identity in spin-isospin 
space, to $\bm{\sigma}(1)\bm{\sigma}(2)$, to
$\bm{\tau}(1)\bm{\tau}(2)$ and to 
$\bm{\sigma}(1)\bm{\sigma}(2)\ \bm{\tau}(1)\bm{\tau}(2)$.
The final expression for these terms reads
\begin{widetext}
\begin{align}
\langle (n_a l_a & j_a \tau_a, n_b l_b j_b \tau_b) J M_J| \bar{V} | (n_c l_c j_c \tau_c, n_d l_d j_d \tau_d) J M_J
\rangle_I = \notag \\
= & \mathcal{N}_I \mathcal{F}_I
\sum_{\Sigma L} i^{-l_a-l_b+l_c+l_d}
\hat{L}^2 \hat{\Sigma}^2 \hat{j}_a \hat{j}_b \hat{j}_c \hat{j}_d
\mathcal{G}_I
\begin{Bmatrix}
l_a & l_b & L \\
\frac{1}{2} & \frac{1}{2} & \Sigma \\
j_a & j_b & J
\end{Bmatrix}
\begin{Bmatrix}
l_c & l_d & L \\
\frac{1}{2} & \frac{1}{2} & \Sigma \\
j_c & j_d & J
\end{Bmatrix}
\notag
\\
&\frac{\lambda^2 \lambda'^2 }{\pi^3} \sum_{\substack{n_i N_i \\ n_f N_f}}
M_{L}(N_f L n_f 0; n_a
l_a n_b
l_b)
M_{L}(N_i L n_i 0; n_c
l_c n_d
l_d) \label{eq:matr_el_cut} \\
&\int \mathrm{d} R R^2
R_{N_f L}(\sqrt{2} \beta R)
g (R)
R_{N_i L}(\sqrt{2} \beta R) \notag \\
&\int \mathrm{d} r \, r R_{n_i 0}(\beta r) j_{1}(r \lambda)
\int \mathrm{d} r' \, r' R_{n_f 0}(\beta r') j_{1}(r' \lambda'). \notag 
\end{align}
\end{widetext}
Here the single-particle states are labelled by the
usual quantum numbers in spherical symmetry, $n, l, j$, 
together with the third component of the isospin $\tau$. 
These single-particle states are expanded in the harmonic
oscillator basis, and $\beta$ is the harmonic oscillator parameter,
$\beta\equiv \sqrt{m\omega/\hbar}$. The harmonic oscillator 
single-particle states and their radial wave functions $R$ are 
defined in Eq. \eqref{eq:spho}. The symbol $M_L$ corresponds to 
the Brody-Moshinsky coefficients. The quantities 
$\mathcal{N}_I,\mathcal{F}_I,\mathcal{G}_I$
are defined in the Appendix. Although the structure of the
formula should look clear, as it includes the transformations (i)
to $LS$-coupling, (ii) to the harmonic oscillator basis, (ii) to
the center-of-mass and relative coordinates, reading the Appendix
may shed some further light on it. 

The two-body matrix elements \eqref{eq:matr_el_cut} constitute 
the backbone of our calculation. Nonetheless, since as 
explained above we will also use the renormalized interaction at the mean field level, it is useful to provide in this
subsection the final form of the one-body matrix elements of the HF Hamiltonian, that are
\begin{align}
h^{(\alpha)}_{ab} & =  \,\, t_{ab} \label{eq:firstkin}\\
&+ \sum_{{\substack{\beta \\ \varepsilon_{\beta} \leq
\varepsilon_{F}}}} \sum_{c d} c_{\beta,c}^* \langle a c | \bar{V} | b d \rangle c_{\beta,d} \label{eq:secondinter} \\
& + \frac{1}{2} \sum_{{\substack{\beta \gamma \\ \varepsilon_{\beta,\gamma} \leq
\varepsilon_{F}}}} \,\, \,\sum_{{cdef}} c_{\beta,c}^* c_{\gamma,d}^* \langle c d |
\frac{\partial{\bar{V}}}{\partial c_{\alpha,a}^*} | ef \rangle c_{\beta,e} c_{\gamma,f} \label{eq:thirdrearr},
\end{align}
where $c$ denotes the expansion coefficients of the s.p. states 
on the harmonic oscillator basis, the 
Greek letters represent the set of quantum number which identify a s.p. state and the Latin letters indicate the harmonic oscillator basis quantum number.

The explicit expression for the term \eqref{eq:firstkin} can be easily found in Ref.~\cite{bertschbook}.
The second term \eqref{eq:secondinter} can be written using Eq.~\eqref{eq:matr_el_cut}.
Nevertheless, the expression can be further simplified because of two simple considerations \cite{Vautherin:1972}:
\begin{itemize}
\item we are dealing with even-even nuclei, thus the matrix elements of the operator $\bm{\sigma}(1)\bm{\sigma}(2)$ vanishes;
\item there is no charge mixing of the HF states, so the isospin exchange operator $P_{\tau}$ reduces to a Kronecker delta.
\end{itemize}
With these simplifications and by using the orthogonality relations for the 9$-j$ symbol, we get
\begin{widetext}
\begin{align}
\sum_{{\substack{\beta \\ \varepsilon_{\beta} \leq
\varepsilon_{F}}}} &\sum_{c d} c_{\beta,c}^* \langle a c | \bar{V} | b d \rangle c_{\beta,d} =
\sum_{{\substack{\beta \\ \varepsilon_{\beta} \leq
\varepsilon_{F}}}} \sum_{c d} \sum_{J} \frac{\hat{J}^2}{\hat{j}^2_{\alpha}} c_{\beta,c}^* \langle (a c) JM | \bar{V} | (b d)JM \rangle c_{\beta,d} \notag \\
=&
\sum_{{\substack{\beta \\ \varepsilon_{\beta} \leq
\varepsilon_{F}}}} \sum_{c d} c_{\beta,c}^* c_{\beta,d} \sum_{L} \sum_{\substack{n_i N_i \\ n_f N_f}}
\frac{\hat{L}^2 \hat{j}^2_{\beta}}{\hat{l}^2_{\alpha} \hat{l}^2_{\beta}}
M_{L}(N_f L n_f 0; a l_{\alpha} c l_{\beta})
M_{L}(N_i L n_i 0; b l_{\alpha} d l_{\beta}) \notag \\
&\qquad
\times \left(1-\frac{1}{2}\delta_{q_\alpha,q_\beta}\right) \int \mathrm{d} R R^2
R_{N_f L}(\sqrt{2} \beta R)
g (R)
R_{N_i L}(\sqrt{2} \beta R) \\
&\qquad \times \frac{\lambda^2 \lambda'^2 }{\pi^3} \int \mathrm{d} r \, r R_{n_i 0}(\beta r) j_{1}(r \lambda)
\int \mathrm{d} r' \, r' R_{n_f 0}(\beta r') j_{1}(r' \lambda'). \notag
\end{align}
\end{widetext}
Following the same strategy, the last term \eqref{eq:thirdrearr}, which is the rearrangement term, can be written as
\begin{widetext}
\begin{align}
\frac{1}{2} \sum_{{\substack{\beta \gamma \\ \varepsilon_{\beta,\gamma} \leq
\varepsilon_{F}}}} \,\, \,&\sum_{\substack{cd \\ ef}} c_{\beta,c}^* c_{\gamma,d}^* \langle c d |
\frac{\partial{\bar{V}}}{\partial c_{\alpha,a}^*} | ef \rangle c_{\beta,e} c_{\gamma,f} \notag \\
=&
\frac{1}{2} \sum_{{\substack{\beta \gamma \\ \varepsilon_{\beta,\gamma} \leq
\varepsilon_{F}}}} \,\, \,\sum_{\substack{cd \\ ef}} c_{\beta,c}^* c_{\gamma,d}^* c_{\beta,e} c_{\gamma,f} \notag \\
& \times \sum_{L} \sum_{\substack{n_i N_i \\ n_f N_f}}
\frac{\hat{L}^2 \hat{j}^2_{\gamma} \hat{j}^2_{\beta}}{\hat{l}^2_{\gamma} \hat{l}^2_{\beta}}
M_{L}(N_f L n_f 0; c l_{\beta} d l_{\gamma})
M_{L}(N_i L n_i 0; e l_{\beta} f l_{\gamma}) \notag \\
&\qquad
\times \left(1-\frac{1}{2}\delta_{q_\gamma,q_\beta}\right) \int \mathrm{d} R R^2
R_{N_f L}(\sqrt{2} \beta R)
g'(R)
R_{N_i L}(\sqrt{2} \beta R) \\
&\qquad \times \frac{\lambda^2 \lambda'^2 }{\pi^3} \int \mathrm{d} r \, r R_{n_i 0}(\beta r) j_{1}(r \lambda)
\int \mathrm{d} r' \, r' R_{n_f 0}(\beta r') j_{1}(r' \lambda'), \notag
\end{align}
\end{widetext}
where $g'(R)=\frac{t_3 \alpha}{24\pi} R_{a l_{\alpha}}(\beta R) R_{b l_{\alpha}}(\beta R) \rho^{\alpha-1}(R)$.

\begin{table*}
\caption{Parameter sets (named SkP$_{\Lambda}$) obtained in the fits associated with different values of the cutoff $\Lambda$ compared with the original set SkP, labeled with SkP (first line) \cite{kassemprivate}. \label{tab:paramSKPKass}}
\centering
\begin{tabular}{lllllllll}
\hline
  & \multicolumn{1}{c}{$t_0$} & \multicolumn{1}{c}{$t_3$} & \multicolumn{1}{c}{$\alpha$} &&
  & \multicolumn{1}{c}{$t_0$} & \multicolumn{1}{c}{$t_3$} & \multicolumn{1}{c}{$\alpha$} \\
\hline
 SkP           & $-$2931.70 & 18709.00 & \multicolumn{1}{c}{$1/6$} & & & & & \\
 SkP$_{0.1}$   & $-$2937.45 & 18758.12 & 0.16674  &\ \ \ \ \ \ \ \ \ \ & SkP$_{1.9}$   &  $-$649.68 &  7431.97 & 1.13340 \\
 SkP$_{0.2}$   & $-$2931.54 & 18723.70 & 0.16713  &\ & SkP$_{2.0}$   &  $-$618.70 &  7062.93 & 1.16305 \\
 SkP$_{0.3}$   & $-$2906.45 & 18577.75 & 0.16881  &\ & SkP$_{2.1}$   &  $-$593.41 &  6596.73 & 1.16744 \\
 SkP$_{0.4}$   & $-$2842.25 & 18204.63 & 0.17328  &\ & SkP$_{2.2}$   &  $-$573.43 &  6052.99 & 1.14457 \\
 SkP$_{0.5}$   & $-$2719.66 & 17494.17 & 0.18249  &\ & SkP$_{2.3}$   &  $-$558.79 &  5469.05 & 1.09369 \\
 SkP$_{0.6}$   & $-$2531.08 & 16406.95 & 0.19873  &\ & SkP$_{2.4}$   &  $-$549.99 &  4892.54 & 1.01547 \\
 SkP$_{0.7}$   & $-$2288.58 & 15022.28 & 0.22432  &\ & SkP$_{2.5}$   &  $-$548.24 &  4374.67 & 0.91252 \\
 SkP$_{0.8}$   & $-$2020.60 & 13517.37 & 0.26140  &\ & SkP$_{3.0}$   &  $-$544.99 &  3624.67 & 0.66267 \\
 SkP$_{0.9}$   & $-$1758.46 & 12085.78 & 0.31144  &\ & SkP$_{3.5}$   &  $-$514.79 &  3386.33 & 0.62361 \\
 SkP$_{1.0}$   & $-$1524.15 & 10862.96 & 0.37503  &\ & SkP$_{4.0}$   &  $-$489.40 &  3180.44 & 0.59654 \\
 SkP$_{1.1}$   & $-$1326.93 &  9904.53 & 0.45153  &\ & SkP$_{5.0}$   &  $-$448.19 &  2858.89 & 0.56329 \\
 SkP$_{1.2}$   & $-$1166.61 &  9204.84 & 0.53904  &\ & SkP$_{8.0}$   &  $-$368.24 &  2279.23 & 0.52259 \\
 SkP$_{1.3}$   & $-$1038.29 &  8724.34 & 0.63454  &\ & SkP$_{10.0}$  &  $-$334.14 &  2045.30 & 0.51106 \\
 SkP$_{1.4}$   &  $-$935.83 &  8409.46 & 0.73409  &\ & SkP$_{20.0}$  &  $-$244.47 &  1457.29 & 0.49046 \\
 SkP$_{1.5}$   &  $-$853.56 &  8203.44 & 0.83327  &\ & SkP$_{40.0}$  &  $-$176.94 &  1035.19 & 0.48119 \\
 SkP$_{1.6}$   &  $-$786.87 &  8050.45 & 0.92736  &\ & SkP$_{60.0}$  &  $-$145.95 &   846.69 & 0.47822 \\
 SkP$_{1.7}$   &  $-$732.24 &  7899.09 & 1.01172  &\ & SkP$_{80.0}$  &  $-$127.16 &   733.92 & 0.47675 \\
 SkP$_{1.8}$   &  $-$687.11 &  7704.42 & 1.08184  &\ & SkP$_{100.0}$ &  $-$114.20 &   656.81 & 0.47589 \\
\hline
\end{tabular}
\end{table*}

%

\section{Results}\label{res}

In our work we have focused on the calculation of the total energy 
(\ref{de-finite-1}) in $^{16}$O. As explained 
in the previous Sections, we aim at using interactions 
fitted with a cutoff regularization in uniform matter
and check that this strategy is enough to prevent the 
divergence of the total energy in the finite system.
The relation between the cutoffs that are used throughout 
our procedure has been given in Eq. (\ref{twocutoffs}) above,
and reads
\begin{displaymath}
\lambda = \sqrt{2}\Lambda.
\end{displaymath}
The interactions $V_\Lambda$ associated with the re-fit of symmetric 
matter,
when the second-order contribution has an associated 
cutoff $\Lambda$, are provided in Table \ref{tab:paramSKPKass}.
As already mentioned, we employ an harmonic oscillator basis. The
oscillator parameter $\beta\equiv \sqrt{m\omega/\hbar}$ is 0.5 fm$^{-1}$ 
and the number of oscillator shells is $n_{\rm max}$ = 10. 
The radial wave functions are calculated up to a maximum value of
$r$ given by $R$ = 12 fm.

\begin{figure}[hbt]
\vspace{1.0cm}
\includegraphics[width=0.9\linewidth]{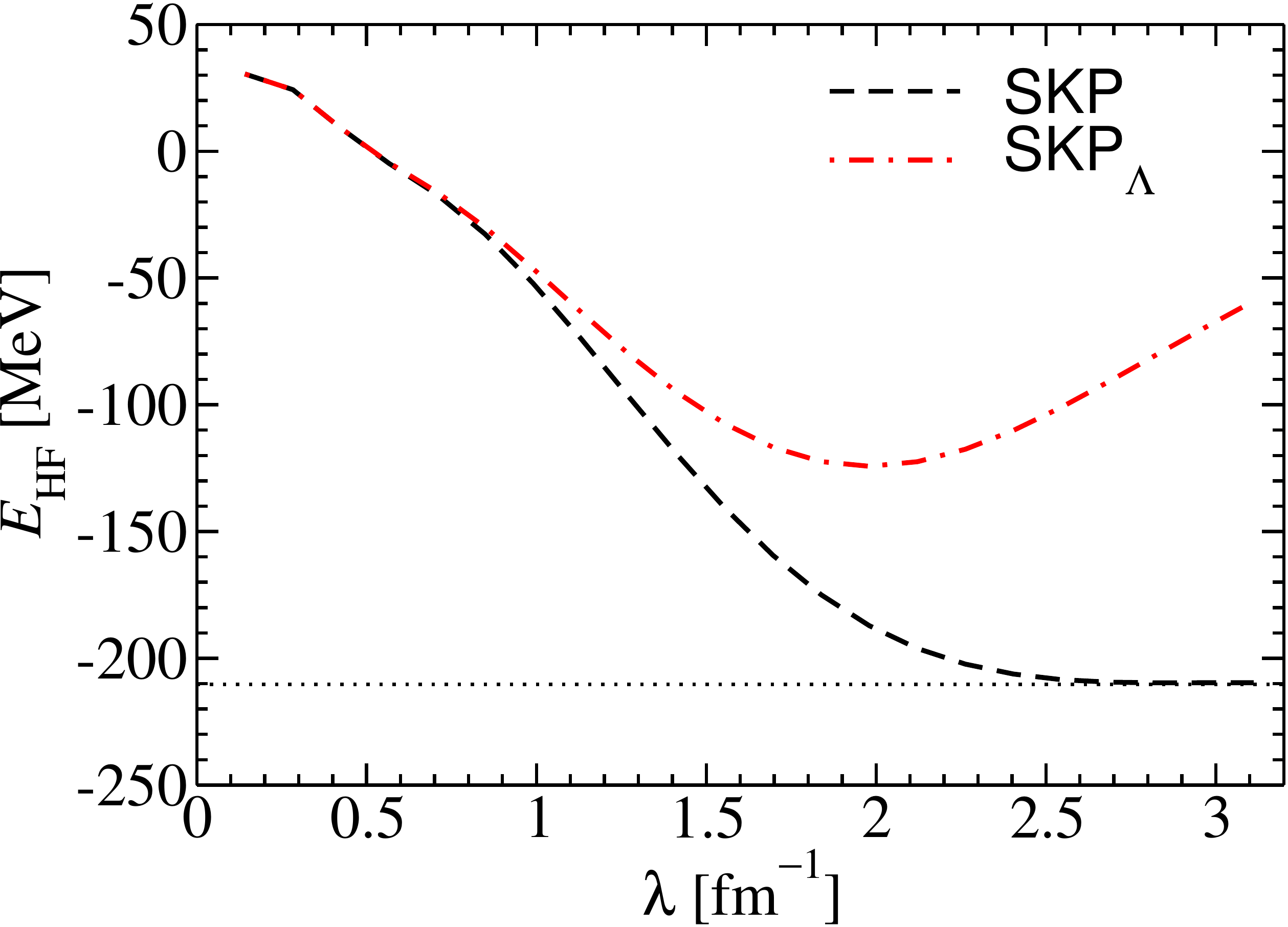}
\caption{(Color online) Total HF energy as a function of a cutoff $\lambda$. The dashed
(dot-dashed) curve corresponds to the result obtained with SkP with only
$t_0$, $t_3$ terms (SkP$_\Lambda$). The thin dotted line corresponds
to the result without any cutoff, obtained with SkP with only $t_0$, $t_3$ term, 
and is meant to guide the eye for the convergence of the dashed line. See the
text for a more detailed discussion.}
\label{fig2}
\end{figure}

\begin{figure}[hbt]
\vspace{1.0cm}
\includegraphics[width=0.9\linewidth]{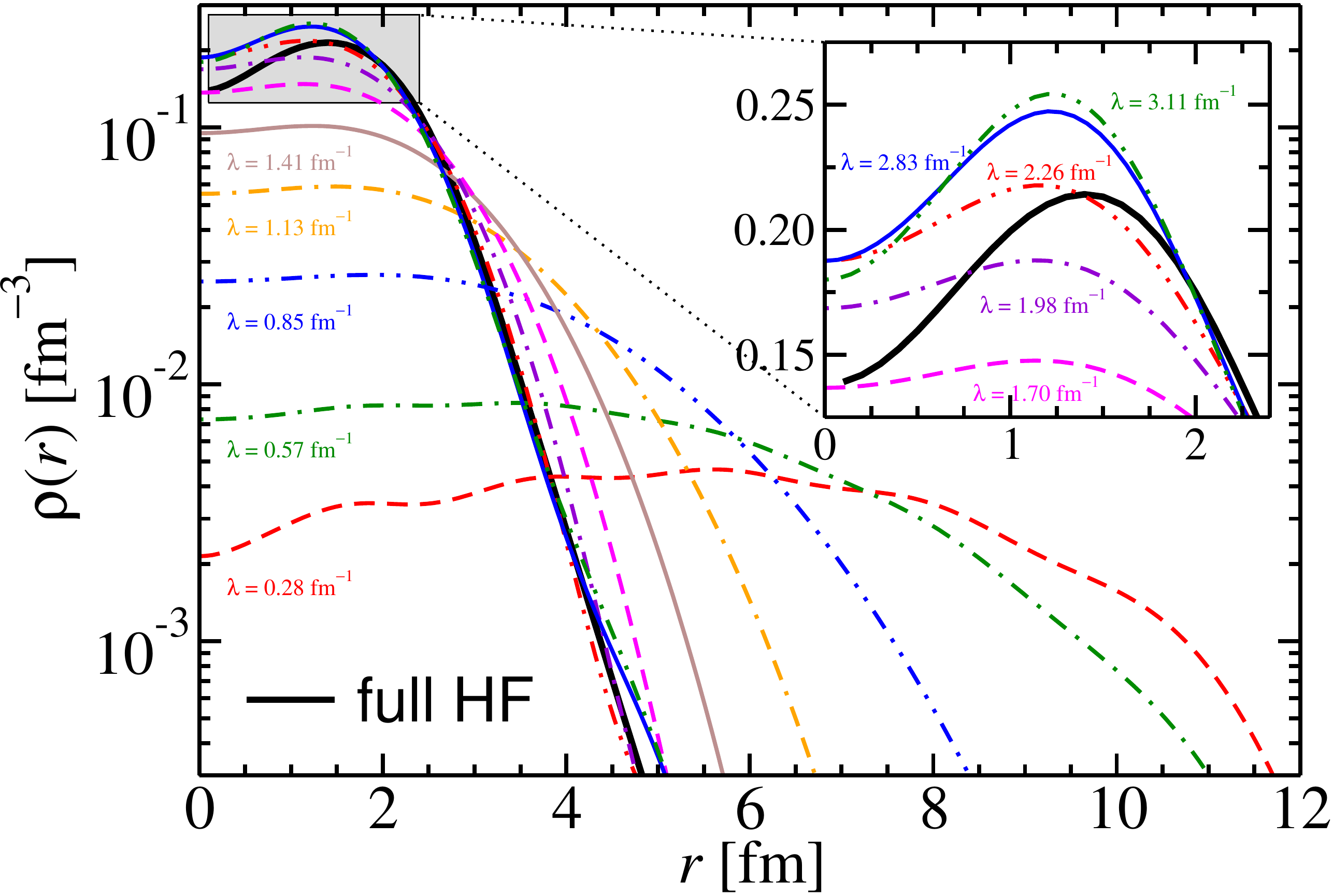}
\caption{(Color online) Total density profiles obtained with the renormalized SkP$_\Lambda$ 
interactions. The thick black line refers to the bare interaction. The inset shows some
detail of the region in which the density attains its largest values.}
\label{fig1}
\end{figure}

In Fig.~\ref{fig2} we display the mean-field energy obtained with the 
renormalized interactions, as a function of $\lambda$, by means of the line labeled with
SkP$_\Lambda$. As a reference we provide the same quantity calculated with the bare interaction
SkP (line labeled with SkP), and the mean field value (without cutoff) which is 
associated with the horizontal line and is 
$-210.3$ MeV. We stress here that the velocity-dependent terms of
the original SkP interaction have been dropped. The original
SkP set, if of course all terms (central terms, both velocity-independent but also
velocity-dependent ones, spin-orbit term and Coulomb term) are retained, 
reasonably reproduces the experimental value of the
binding energy of $^{16}$O which is known to be $-127.619$ MeV 
\cite{Audi}.
The HF result obtained with SkP can be understood, since when the 
cutoff increases
the calculation tends obviously to the exact one, while reducing 
the momentum components amounts to trying to minimize the energy 
in a Hilbert space which is not complete: although the variational
principle cannot be rigorously invoked, it is plausible that
the energy does not attain
its minimum and is instead smaller in absolute value. It is
instructive to note that the convergence to the exact result
is obtained only for $\lambda$ of the order of 2.5 fm$^{-1}$: in fact,
inside this light nucleus the density can rise up to 0.25 fm$^{-3}$ (cf.
Fig. \ref{fig1}) and the associated maximum effective (local) $k_F$ is 
therefore $\approx$ 1.54 fm$^{-1}$, so that the maximum value for
the momenta defined in Eq. (\ref{eq:firstkk}) can be as high as 2.2 
fm$^{-1}$. 
The result associated with the
renormalized interaction is more subtle to understand.  For low values of the cutoff \sout{energy}
the renormalization of the interaction is not significant (as one can notice from the values
of the parameters in Table \ref{tab:paramSKPKass}). 
This is understandable, since for small values of $\lambda$ the second-order
contribution in infinite matter is small, and one needs to weakly renormalize  
the interaction in order to obtain in the same system the HF energy associated with the 
bare interaction. As a consequence, for small values of $\lambda$ the curves associated with the bare and renormalized
interactions overlap. However, for large values of $\lambda$ the total energy still decreases in
absolute value when it is calculated with the renormalized 
interaction. In this case, in fact, most of the momentum
components are retained, but the interaction is strongly renormalized (again, this
can be seen from the values of the parameters in Table \ref{tab:paramSKPKass}) and, as 
a consequence, the mean-field total energy is small. 
As a conclusion, we infer from Fig. \ref{fig2} that for either too small or too large values of
$\lambda$ the system calculated at mean-field level with the renormalized interaction is
far away from the system we would like to reproduce by adding the second-order contribution:
in other words, perturbation theory is doomed to fail for those values of $\lambda$, especially if we start from a situation in which 
the total energy is positive at mean-field but not only in that case. In practice, 
we restrict our following discussion to, and draw conclusions from, 
values of $\lambda$ between $\approx$ 2 and 2.7 
fm$^{-1}$ (although we will show in the figures some results associated with a broader
range of values for $\lambda$). 

\begin{figure*}[hbt]
\vspace{1.0cm}
\includegraphics[width=0.45\textwidth]{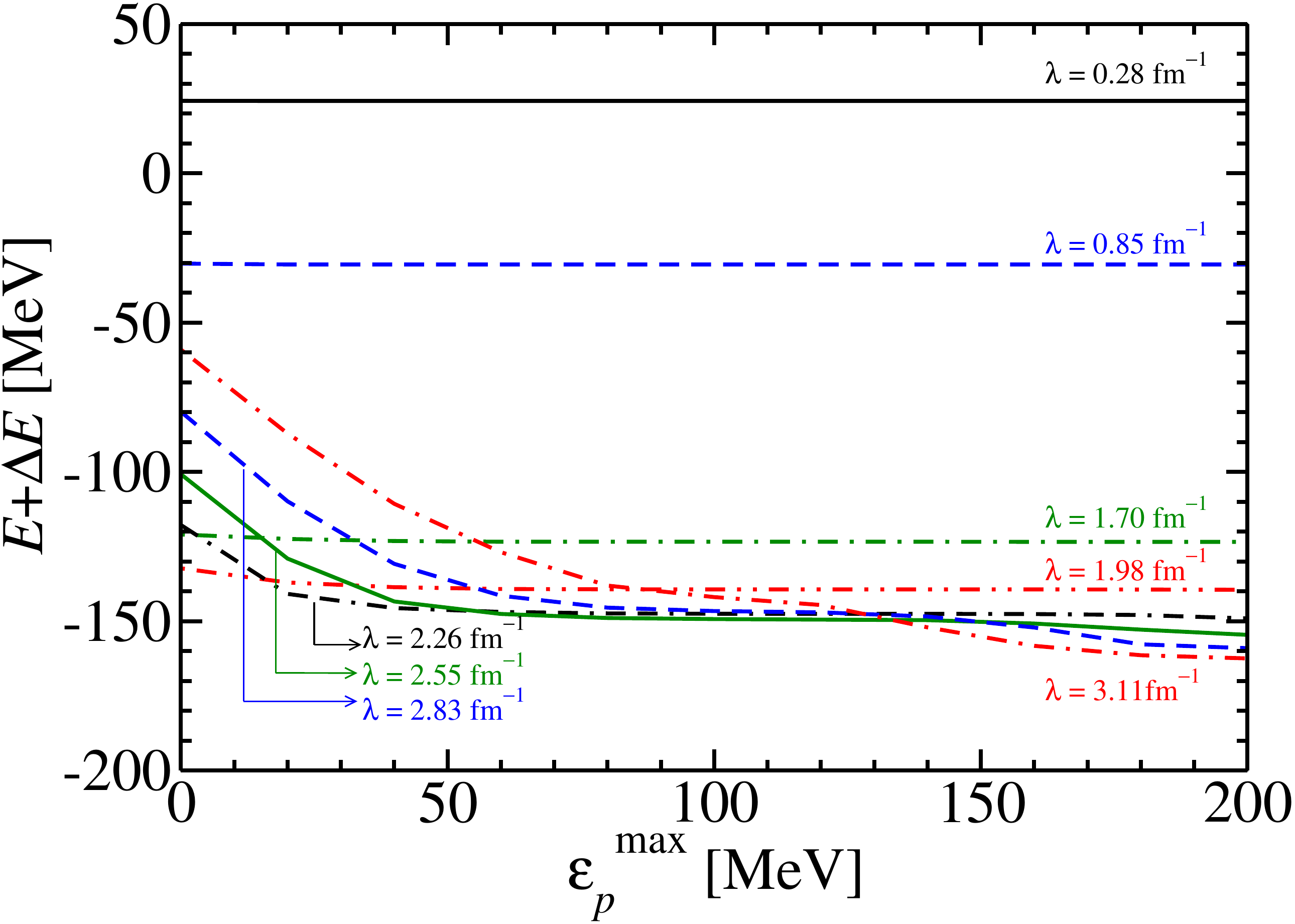}
\hfill
\includegraphics[width=0.45\textwidth]{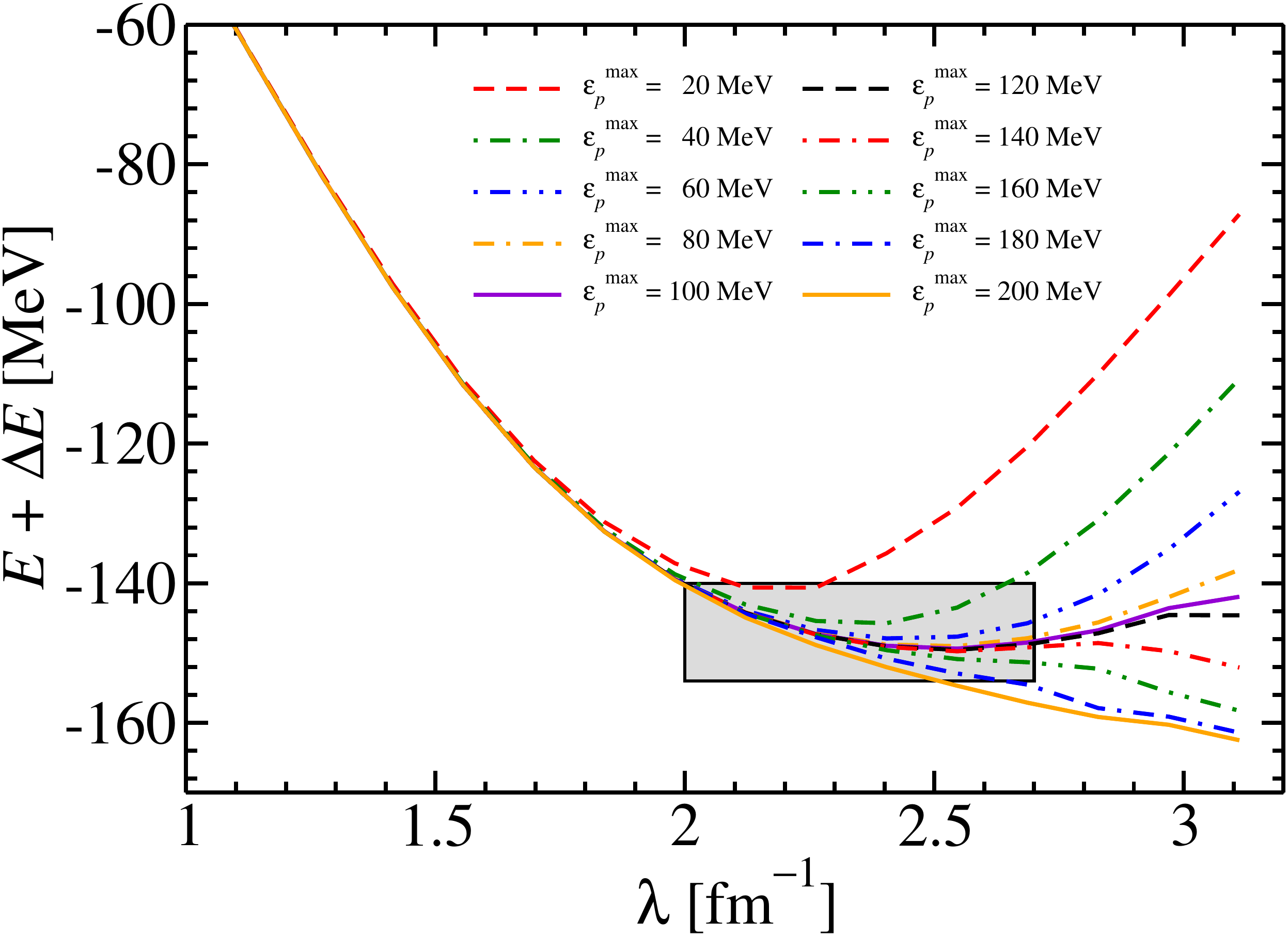}
\caption{(Color online) Total energy at second order as a function of the maximum particle energy (left
panel) or as a function of the cutoff (right panel). All curves are obtained with renormalized 
interactions. See the text for a discussion.}
\label{fig3}
\end{figure*}

This view is in part confirmed by the results shown in Fig. \ref{fig1}: here we display the different
profiles for the total density emerging from the HF calculations when different renormalized
interactions are employed. Along the same line of the discussion in the previous paragraph,
if the cutoff is small a large fraction of the high momentum or small 
distance components of the relative motion are cut, and the system becomes 
very dilute, almost like a uniform unbound nucleon gas. 

\begin{figure}[hbt]
\includegraphics[width=\linewidth]{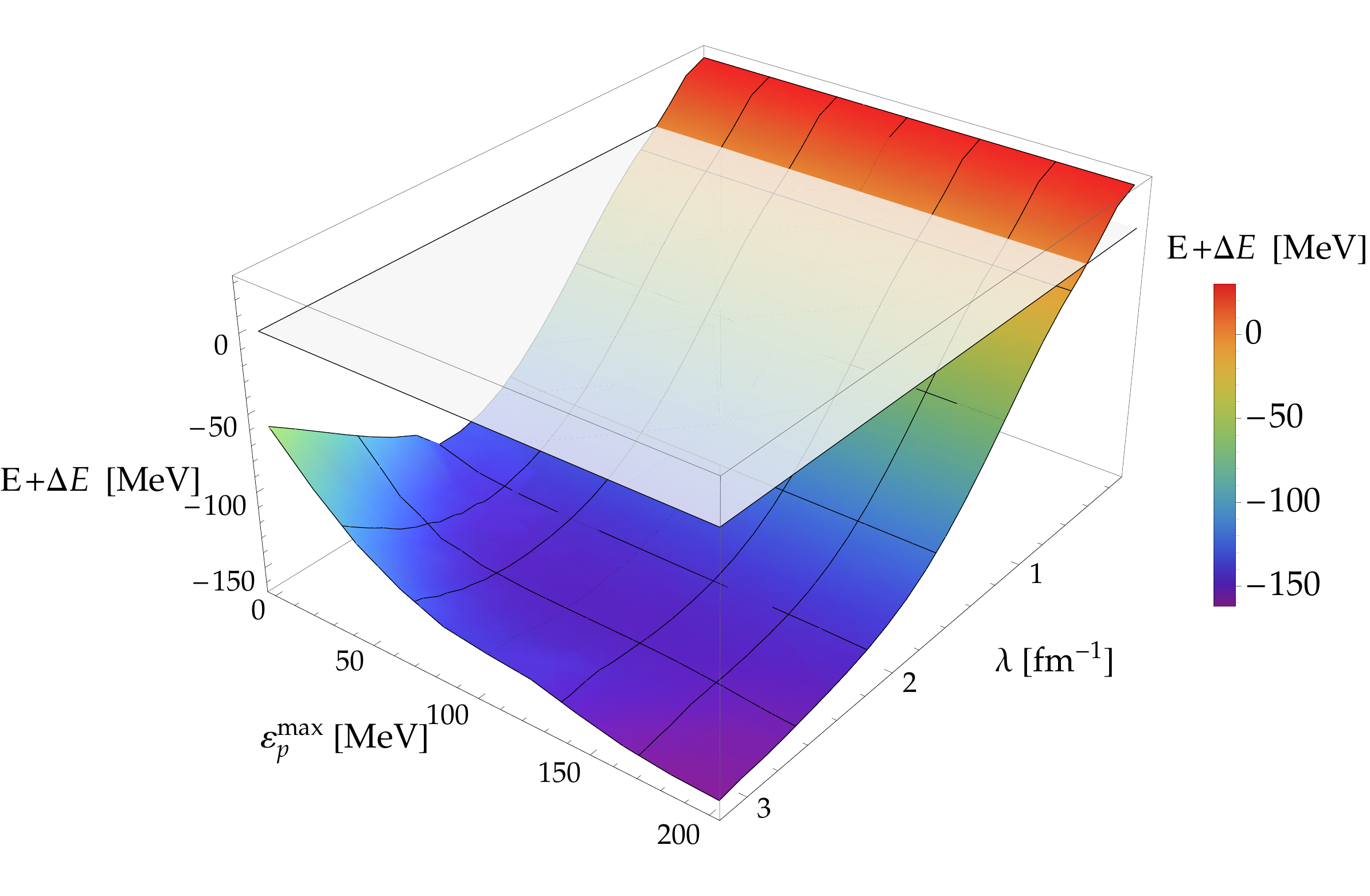}
\caption{(Color online) The same as Fig.~\ref{fig3} in a three-dimensional representation, that is, total energy
at second order as a function of both the maximum particle energy and the cutoff.}
\label{fig4}
\end{figure}

We now discuss our main results, that are summarized in the three panels of Figs.~\ref{fig3} and \ref{fig4}. In the left panel of Fig.~\ref{fig3}, the total energy calculated at second
order with the renormalized interactions is shown, for various values of $\lambda$, as a 
function of the maximum particle energy $\varepsilon_p^{\rm max}$. For the sake of clarity only
a selection of the results obtained with the interactions associated with different values
of $\Lambda$ are displayed. For values of $\lambda$ between $\approx$ 2 and 2.7 
fm$^{-1}$, the results are close to one another. Even more importantly, these results 
do not depend on the value of $\varepsilon_p^{\rm max}$, at least if this is larger
than $\approx$ 80 MeV. This can be understood on the basis of a simple semiclassical
argument: particles having energies larger than $\approx$ 80 MeV, and having thus
very large kinetic energies, would contribute to the total energy through matrix elements
associated with momentum components that are actually eliminated by our choice
of the cutoff. 
Therefore, the most important conclusion is that our proposed renormalization 
procedure can work, and the extra scale associated with the maximum value of the
particle energy, or with the total momentum, does not spoil that procedure.

The stability of the renormalized results for the second order energy, is also visible
in the right panel of Fig.~\ref{fig3}. The curves associated with values of $\lambda$ 
between $\approx$ 2 and 2.7 fm$^{-1}$, for a broad range of values of $\varepsilon_p^{\rm max}$,
lie in the shaded box that corresponds to $\approx$ 10\% error in the total energy.
A quick glance of the behavior of our results is allowed by the plot of Fig.~\ref{fig4},
that collects the same information already provided in the two panels of 
Fig.~\ref{fig3} by means of
a more intuitive three-dimensional global representation.  

%

\section{Conclusions}\label{concl}

The problem associated with the fact that zero-range forces produce
divergent results when employed beyond mean-field has become object of 
a renewed interest. Skyrme forces have zero-range character but also
Gogny forces possess a zero-range terms; pure finite-range forces have not
been so widely developed and systematically applied in non-relativistic approaches.

The problem of the renormalization of these divergences has been tackled
first in a simple system like uniform nuclear matter, restricting to the case
of the second-order correction to the energy. The purpose of this work is
to make a significant first step in the direction of a full renormalization scheme
for the Skyrme force in finite nuclei. Our main idea is to work in a harmonic
oscillator basis, so that the center-of mass and relative motion coordinates
and associated momenta can be neatly separated. The cutoff we set on the
momentum associated with the relative motion can be then related to the
one used in the previous calculations of nuclear matter. We have illustrated
such formal scheme in full detail in this paper.

As a numerical application, we have limited ourselves to $^{16}$O calculated
with a simple, momentum-independent Skyrme force. Our main results are listed as follows: 
\begin{itemize}
\item under certain conditions, it is possible to relate the cutoff $\lambda$ on the relative 
momentum to the cutoff $\Lambda$ that has been previously introduced in \cite{Mog:1,Mog:2}
[cf. Eq. (\ref{twocutoffs}) and the related discussion];
\item if calculations of the total energy at second order are envisaged, with a 
cutoff $\lambda$, the interactions introduced in uniform matter using the
associated value of $\Lambda$ can be employed;
\item the practical way to do these calculations is to work in harmonic oscillator
basis and change the form of the interaction so that relative-momentum components
larger than $\lambda$ are cut;
\item at least for a reasonable range of values, $\lambda \approx 2 - 2.7$ fm${}^{-1}$, the results turn out
to be stable, namely the divergence does not show up;
\item in particular, the second energy scale associated with the total momentum, 
that in finite systems is associated with the maximum value of the particle energy
$\varepsilon_p^{\rm max}$, does not seem to spoil this stability. This 
can be justified
by semiclassical arguments.
\end{itemize}

In terms of perspectives, several issues remain to be considered. Our results
look promising only in a limited window for values of the cutoff. We are
inclined to think that this is due to the choice of a perturbative scheme to
calculate the second-order energies. At variance with the case of infinite
matter, a consistent second-order calculation that employs the proper
equations (Dyson equation for the s.p. energies, Koltun sum rule for
the total energy) is probably called for. On top of this, we are dealing with
a simple Skyrme force which is not very realistic and the extension to
the full force must be also envisaged. Only after this, one could judge if
the plan of devising a zero-range force that is fitted and consistently 
used beyond mean-field, is feasible. In this respect, our results can be
seen as promising but we can draw only qualitative, and not too much
quantitative, conclusions at the present stage.

%

\section*{Acknowledgments}   
M.B. thanks IPN Orsay for the warm hospitality during the time when part of this work was 
carried out. 
The authors are grateful to P. F. Bortignon, B. G. Carlsson, M. Grasso, E. Khan, K. Moghrabi, 
J. Piekarewicz, P. Ring and N. Van Giai for useful discussions and comments.

%

\appendix

\section{Derivation of the pp-coupled matrix elements}

We discuss here, in some detail, the steps that are necessary
to evaluate the particle-particle coupled matrix elements of 
Eq. (\ref{eq:ppcouplme}) and we give some intermediate formula
which can be useful for the reader.  

Let us consider a two-particle state 
$\vert n_a l_a j_a m_a, n_b l_b j_b m_b \rangle$ in a harmonic oscillator potential. The 
single particle wave functions are 
\begin{equation}\label{eq:spho}
\vert n l j m \tau \rangle = \psi^{\tau}_{nljm} (\bm{r}) = i^l R_{nl}(\beta r) \left[Y_l(\hat{r}) \otimes \chi_{\frac{1}{2}}
\right]_{jm} \xi_{\tau},
\end{equation}
where $\beta^2 = \frac{m \omega}{\hbar}$.
If the two states are coupled to total angular momentum $JM$ as 
in Eq. (\ref{eq:ppcouplme}), we need to switch from the $j$--$j$ coupling 
scheme to the $L$--$S$ one before making the Brody-Moshinsky transformation. Then, the two-particle states read
\begin{widetext}
\begin{align}
| n_a & l_a j_a m_a \tau_a, n_b l_b j_b m_b \tau_b \rangle =
\begin{aligned}[t]
\sum_{\substack{J M_J \\ \Lambda \Sigma}} \hat{\Lambda} \hat{\Sigma} \hat{j}_a
\hat{j}_b & \langle j_a m_a j_b m_b | J M_J \rangle
\begin{Bmatrix}
l_a & l_b & \Lambda \\
\frac{1}{2} & \frac{1}{2} & \Sigma \\
j_a & j_b & J
\end{Bmatrix}
| [ n_a n_b, (l_a , l_b) \Lambda, (\frac{1}{2},\frac{1}{2}) \Sigma, \tau_a \tau_b ] J M_J \rangle
\end{aligned}
\notag \\
&\begin{aligned}
=&\sum_{J \Lambda \Sigma} \sum_{\substack{M_\Lambda M_\Sigma M_J \\ M_l M_L}} \sum_{\sigma_a \sigma_b} i^{l_a+l_b}
(-)^{j_a-j_b+\Lambda+M_\Lambda+\Sigma+M_\Sigma+L+l}
\hat{J}^2 \hat{\Lambda}^2 \hat{\Sigma}^2 \hat{j}_a \hat{j}_b
\begin{Bmatrix}
l_a & l_b & \Lambda \\
\frac{1}{2} & \frac{1}{2} & \Sigma \\
j_a & j_b & J
\end{Bmatrix} \\
&
\begin{pmatrix}
j_a & j_b & J \\
m_a & m_b &-M_J
\end{pmatrix}
\begin{pmatrix}
\Lambda & \Sigma & J \\
M_\Lambda & M_\Sigma &-M_J
\end{pmatrix}
\begin{pmatrix}
\frac{1}{2} & \frac{1}{2} & \Sigma \\
\sigma_a & \sigma_b &-M_\Sigma
\end{pmatrix}
\begin{pmatrix}
L & l & \Lambda \\
M_L & M_l &-M_\Lambda
\end{pmatrix}
\\
& \sum_{n l N L} M_{\Lambda}(N L n l; n_a l_a n_b l_b) R_{n l}(\beta r)R_{N L}(\beta R)     Y_{l M_l} (\hat{r}) Y_{L
M_L}(\hat{R}) \chi_{\sigma_a}(1) \chi_{\sigma_b}(2) \xi_{\tau_a}(1) \xi_{\tau_b}(2),
\end{aligned} \label{eq:twostate}
\end{align}
\end{widetext}
where the center of mass and relative motion coordinates have been
introduced as in Eq. (\ref{eq:transIordrR}) and the 
corresponding Brody-Moshinsky coefficients are denoted by 
$M_{\Lambda}$ \cite{lawson,buck96,kamun01}. In addition,
several intermediate quantum numbers are introduced. 

We want to compute the matrix elements of the Skyrme interaction, written as in Eq.~\eqref{eq:interr1r2}, between the two-particle 
states \eqref{eq:twostate}. 
We are interested in the antisymmetrized interaction $\bar{V}=V(1-P_M P_\sigma P_\tau)$, where $P_M$ is the Majorana exchange 
operator while $P_\sigma$ and $P_\tau$ are the spin and isospin exchange operators. 
In the case of the center of mass and relative motion coordinate system the Majorana operator is non-trivial. 
The exchange operator acts on the two-particle state \eqref{eq:twostate} in the following way:
\begin{widetext}
\begin{align}
| n_b & l_b j_b m_b \tau_b, n_a l_a j_a m_a \tau_a \rangle =  P_M P_{\sigma} P_{\tau} | n_a l_a j_a m_a \tau_a, n_b l_b j_b
m_b \tau_b \rangle \notag \\
=&
i^{l_a+l_b} \hat{j}_a \hat{j}_b
\sum_{J M_J} \sum_{\Lambda M_{\Lambda}} \sum_{\Sigma M_{\Sigma}} \sum_{L M_L}  \sum_{l M_l} \sum_{N n} \sum_{\sigma_a
\sigma_b}
(-1)^{l_a+l_b-L}
\hat{J}^2 \hat{\Lambda}^2 \hat{\Sigma}^2
\begin{Bmatrix}
l_a & l_b & \Lambda \\
\frac{1}{2} & \frac{1}{2} & \Sigma \\
j_a & j_b & J
\end{Bmatrix} \notag \\
&
(-1)^{j_a-j_b+\Lambda+M_{\Lambda}+\Sigma+M_{\Sigma}+L+l}
M_{\Lambda}(N L n l; n_a l_a n_b l_b) \notag \\
&R_{n l}(\beta r) R_{N L}(\beta R)
Y_{l M_l} (\hat{r}) Y_{L M_L} (\hat{R}) \notag \\
&
\begin{pmatrix}
j_a & j_b & J   \\
m_a & m_b &-M_J
\end{pmatrix}
\begin{pmatrix}
\Lambda   & \Sigma   & J   \\
M_\Lambda & M_\Sigma &-M_J
\end{pmatrix}
\begin{pmatrix}
\frac{1}{2} & \frac{1}{2} & \Sigma     \\
\sigma_a    & \sigma_b    &-M_{\Sigma}
\end{pmatrix}
\begin{pmatrix}
L   & l   & \Lambda    \\
M_L & M_l &-M_{\Lambda}
\end{pmatrix} \notag \\
& P_{\sigma}\left[\chi_{\sigma_a}(1) \chi_{\sigma_b}(2)\right] P_{\tau} \left[ \xi_{\tau_a}(1) \xi_{\tau_b}(2) \right]\notag,
\end{align}
\end{widetext}
that is equal to Eq.~\eqref{eq:twostate} except for the $P_{\sigma}$, $P_{\tau}$ operators and  the phase factor $(-1)^{l_a+l_b-L}$.
This can be checked by direct calculation. 

As mentioned in the main text (cf. Subsec. \ref{2c}) we provide
separately the expressions for the 
different spin and isospin terms of the pp-coupled
matrix elements~\eqref{eq:ppcouplme}, and we shall use for them, 
respectively, the label $I$ that can assume the values 
$I=0, \sigma, \tau, \sigma \tau$. We also introduce the following quantities:
\begin{widetext}
\begin{equation}\label{eq:cases}
\begin{split}
\mathcal{F}_I &=
\begin{cases}
\delta_{\tau_a \tau_c} \delta_{\tau_b \tau_d} & \text{if } I = 0,\sigma \nonumber \\
\sum_{\mu} (-)^{1+\tau_a+\tau_b+\mu}
\begin{pmatrix}
\frac{1}{2} & 1  & \frac{1}{2} \\
\tau_c   &\mu & -\tau_a
\end{pmatrix}
\begin{pmatrix}
\frac{1}{2} & 1  & \frac{1}{2} \\
\tau_d   &-\mu & -\tau_b
\end{pmatrix} & \text{if } I = \tau,\sigma\tau
\end{cases} \nonumber \\
\mathcal{G}_I &=
\begin{cases}
1 & \text{if } I = 0,\tau \nonumber \\
(-)^{1+\Sigma}
\begin{Bmatrix}
\frac{1}{2} & \frac{1}{2} & \Sigma \\
\frac{1}{2} & \frac{1}{2} & 1
\end{Bmatrix} & \text{if } I = \sigma,\tau
\end{cases} \nonumber \\
\mathcal{N}_I&=
\begin{cases}
\frac{3}{4} & \text{if } I = 0 \nonumber \\
-\frac{3}{2} & \text{if } I = \sigma,\tau \nonumber \\
-9 & \text{if } I = \sigma\tau
\end{cases} \nonumber \\
\mathcal{M}_I&=
\begin{cases}
\left(1-\frac{1}{4} (-1)^{l_c+l_d-L} \right) 
& \text{if } I = 0 \nonumber \\
(-1)^{l_c+l_d-L} & \text{if } 
I = \sigma\tau. 
\end{cases}
\end{split}
\end{equation}

Then, the general expression for the four terms of the matrix elements reads
\begin{align}
\begin{split}
\langle (n_a l_a & j_a \tau_a, n_b l_b j_b \tau_b) J M_J| \bar{V} | (n_c l_c j_c \tau_c, n_d l_d j_d \tau_d) J M_J
\rangle_I =\\
=& \mathcal{N}_I \mathcal{F}_I
\sum_{\substack{\Lambda  \Sigma \\ L l}} i^{-l_a-l_b+l_c+l_d} (-)^l 
\mathcal{M}_I
\mathcal{G}_I
\begin{Bmatrix}
l_a & l_b & \Lambda \\
\frac{1}{2} & \frac{1}{2} & \Sigma \\
j_a & j_b & J
\end{Bmatrix}
\begin{Bmatrix}
l_c & l_d & \Lambda \\
\frac{1}{2} & \frac{1}{2} & \Sigma \\
j_c & j_d & J
\end{Bmatrix} \\
&\frac{\hat{\Lambda}^2 \hat{\Sigma}^2 \hat{j}_a \hat{j}_b \hat{j}_c \hat{j}_d}{\hat{l}}
\sum_{\substack{n_i N_i \\ n_f N_f}}
M_{\Lambda}(N_f L n_f l; n_a
l_a n_b
l_b)
M_{\Lambda}(N_i L n_i l; n_c
l_c n_d
l_d) \\
&\int \mathrm{d} R R^2
R_{N_f L}(\sqrt{2} \beta R)
g (R)
R_{N_i L}(\sqrt{2} \beta R) \\
& \int \mathrm{d} r \mathrm{d} r' \, r^2 r'^2 R_{n_f l}(\beta r')  v_{lm}(r',r)
R_{n_i l}(\beta r).
\end{split}
\label{eq:megenerredcom}
\end{align}
\end{widetext}

It can be a useful exercise to insert in this expression the standard coefficients of the multipole expansion
of the velocity-independent part of the Skyrme interaction, that are
\begin{equation}
\begin{split}
v_{lm}(r',r) =
\frac{(-)^l \hat{l}}{4 \pi} \frac{\delta (r)}{r^2} \frac{\delta (r')}{r'^2}.
\end{split}
\end{equation}
Then the matrix element reads
\begin{widetext}
\begin{align}
\langle (n_a l_a & j_a \tau_a, n_b l_b j_b \tau_b) J M_J| \bar{V} | (n_c l_c j_c \tau_c, n_d l_d j_d \tau_d) J M_J
\rangle_I = \notag \\
=& \mathcal{N}_I \mathcal{F}_I
\sum_{\Sigma L } i^{-l_a-l_b+l_c+l_d}
\frac{\hat{L}^2 \hat{\Sigma}^2 \hat{j}_a \hat{j}_b \hat{j}_c \hat{j}_d}{4 \pi}
\mathcal{G}_I
\begin{Bmatrix}
l_a & l_b & L \\
\frac{1}{2} & \frac{1}{2} & \Sigma \\
j_a & j_b & J
\end{Bmatrix}
\begin{Bmatrix}
l_c & l_d & L \\
\frac{1}{2} & \frac{1}{2} & \Sigma \\
j_c & j_d & J
\end{Bmatrix} \notag \\
&\sum_{\substack{n_i N_i \\ n_f N_f}}
M_{L}(N_f L n_f 0; n_a
l_a n_b
l_b)
M_{L}(N_i L n_i 0; n_c
l_c n_d
l_d) \label{eq:matrelem_skyrme} \\
& R_{n_i 0}(0) R_{n_f 0}(0) \int \mathrm{d} R R^2
R_{N_f L}(\sqrt{2} \beta R)
g (R)
R_{N_i L}(\sqrt{2} \beta R). \notag
\end{align}
\end{widetext}
This expression does not include isospin coupling (when this is considered, cf. the 
analogous expression in e.g. Ref.  \cite{lawson}).

For the renormalized interaction the multipole expansion coefficients can be found, instead, as
\begin{widetext}
\begin{equation}
\begin{split}
v_{lm}^{\lambda\lambda'}(r',r) = & \frac{1}{4 \pi^4} \frac{\lambda^2\lambda'^2}{r r'} j_1(\lambda r)
j_1(\lambda'
r') \frac{(-)^l}{\hat{l}} \sum_m \int
\mathrm{d} \hat{r}' \mathrm{d} \hat{r} \, Y^*_{lm}(\hat{r})Y_{lm}(\hat{r}') \\
= & \frac{1}{4 \pi^4} \frac{\lambda^2\lambda'^2}{r r'} j_1(\lambda r) j_1(\lambda'
r') \frac{(-)^l}{\hat{l}} \sum_m 4\pi \delta_{l,0} \delta_{m,0} 
= \frac{1}{4 \pi^3} \frac{\lambda^2\lambda'^2}{r r'} j_1(\lambda r) 
j_1(\lambda' r') \delta_{l,0};
\end{split}
\end{equation}
then, the corresponding matrix element reads
\begin{align}
\langle (n_a l_a & j_a \tau_a, n_b l_b j_b \tau_b) J M_J| \bar{V} | (n_c l_c j_c \tau_c, n_d l_d j_d \tau_d) J M_J
\rangle_I = \notag \\
= & \mathcal{N}_I \mathcal{F}_I
\sum_{\Sigma L} i^{-l_a-l_b+l_c+l_d}
\hat{L}^2 \hat{\Sigma}^2 \hat{j}_a \hat{j}_b \hat{j}_c \hat{j}_d
\mathcal{G}_I
\begin{Bmatrix}
l_a & l_b & L \\
\frac{1}{2} & \frac{1}{2} & \Sigma \\
j_a & j_b & J
\end{Bmatrix}
\begin{Bmatrix}
l_c & l_d & L \\
\frac{1}{2} & \frac{1}{2} & \Sigma \\
j_c & j_d & J
\end{Bmatrix}
\notag
\\
&\frac{\lambda^2 \lambda'^2 }{\pi^3} \sum_{\substack{n_i N_i \\ n_f N_f}}
M_{L}(N_f L n_f 0; n_a
l_a n_b
l_b)
M_{L}(N_i L n_i 0; n_c
l_c n_d
l_d) \notag \\
&\int \mathrm{d} R R^2
R_{N_f L}(\sqrt{2} \beta R)
g (R)
R_{N_i L}(\sqrt{2} \beta R) \notag \\
&\int \mathrm{d} r \, r R_{n_i 0}(\beta r) j_{1}(r \lambda)
\int \mathrm{d} r' \, r' R_{n_f 0}(\beta r') j_{1}(r' \lambda'). \notag 
\end{align}
\end{widetext}
These are the matrix elements displayed in Eq. 
\eqref{eq:matr_el_cut} in the main text. 

%
\bibliography{bibliography}

\end{document}